\documentclass[aps,prl,reprint,superscriptaddress,floatfix]{revtex4-2}

\usepackage{amsmath,amssymb}
\usepackage{graphicx}
\usepackage{bm,mathrsfs}
\usepackage{physics}
\usepackage{hyperref}
\usepackage{placeins}
\graphicspath{{./}{./sm/}{./figure/}{./sm/fig/}}
\hypersetup{
	colorlinks=true,
	linkcolor=blue,
	citecolor=blue,
	urlcolor=blue
}
\newcommand{\parhead}[1]{\noindent\textit{#1.---}}
\begin{document}
	
	\title{Quantum-geometry stabilization of dilute fractional Chern insulators}
	
	\author{Ying-Xing Ding}
	\thanks{These authors contributed equally to this work.}
	\affiliation{
		Beijing National Laboratory for Condensed Matter Physics, Institute of Physics, \\
		Chinese Academy of Sciences, Beijing 100190, China
	}
	\affiliation{
		School of Physical Sciences, \\
		University of Chinese Academy of Sciences, Beijing 100049, China
	}
	
	\author{Li-Min Zhang}
	\thanks{These authors contributed equally to this work.}
	\affiliation{
		CAS Key Laboratory of Quantum Information, \\
		University of Science and Technology of China, Hefei 230026, China
	}
	
	\author{Wen-Tong Li}
	\affiliation{
		Beijing National Laboratory for Condensed Matter Physics, Institute of Physics, \\
		Chinese Academy of Sciences, Beijing 100190, China
	}
	\affiliation{
		School of Physical Sciences, \\
		University of Chinese Academy of Sciences, Beijing 100049, China
	}
	
	\author{D. L. Zhou}
	\affiliation{
		Beijing National Laboratory for Condensed Matter Physics, Institute of Physics, \\
		Chinese Academy of Sciences, Beijing 100190, China
	}
	\affiliation{
		School of Physical Sciences, \\
		University of Chinese Academy of Sciences, Beijing 100049, China
	}
	
	\author{Wu-Ming Liu}
	\email{wliu@iphy.ac.cn}
	\affiliation{
		Beijing National Laboratory for Condensed Matter Physics, Institute of Physics, \\
		Chinese Academy of Sciences, Beijing 100190, China
	}

	\begin{abstract}
		Fractional Chern insulators have attracted broad interest as lattice analogs of fractional quantum Hall states without Landau levels. However, low-filling fractional Chern insulators are fragile because charge-ordered phases can compete strongly with the fractional topological liquid.
		Here, we propose a center-decorated kagome model, motivated by geometry-tunable artificial lattices, in which the center-site hopping $t_2$ provides a direct knob for the quantum geometry of an isolated $C=1$ flat band.
		Here quantum geometry refers to the Berry curvature and Fubini--Study metric, which determine the form factors of interactions projected into the Chern band. Exact diagonalization shows that tuning $t_2$ away from the flatness-optimized kagome limit reduces the trace-condition deviation, suppresses competing charge order, and enhances the many-body stability at both $\nu=1/3$ and the more fragile $\nu=1/5$ filling.
		At $\nu=1/5$, this stability-enhanced window persists under nearby interaction profiles, including variations of the dominant third-neighbor repulsion and weak nearest-neighbor admixtures.
		Low-energy spectra, spectral flow, quasihole and entanglement counting, static structure factors, and the quantized total many-body Chern number $C_{\mathrm{tot}}=1$ consistently support Laughlin-like fractional Chern insulators.
		These results identify quantum-geometry engineering as a route to stabilizing dilute fractional Chern insulators beyond band-flatness optimization alone.
	\end{abstract}
	
	\maketitle
	
	
	\parhead{Introduction} Fractional Chern insulators (FCIs) realize fractional quantum Hall physics in partially filled Chern bands without Landau levels~\cite{PhysRevLett.106.236802,PhysRevLett.106.236803,PhysRevLett.106.236804,WOS:000294805300016,PhysRevB.88.035101,PhysRevX.14.041040}.
	They support fractionalized excitations, topological ground-state degeneracy, and quantized Hall response in lattice systems where topology is generated by Berry curvature rather than by an external magnetic field.
	Recent moir\'e experiments have made such fractional Chern physics experimentally relevant~\cite{PhysRevLett.133.066601,PhysRevLett.133.186602,PhysRevLett.133.206502,PhysRevLett.133.206504,PhysRevB.108.085117,PhysRevB.109.045147,PhysRevB.109.205121,PhysRevB.109.205122,WOS:001537392100025,WOS:001620290500001,WOS:001361300200018,WOS:000730754700026,WOS:001610291300001,WOS:001078346100001,WOS:000733421800008,WOS:001127160700001}. Programmable optical lattices offer a complementary platform with microscopic control over topological flat bands and fractional responses~\cite{PhysRevLett.94.086803,PhysRevA.76.023613,PhysRevLett.125.236401,Nature.619.495}. The central challenge is to stabilize low-filling fractional liquids against lattice-induced competing orders.
	
	While $\nu=1/3$ FCIs are comparatively robust, lower fillings such as $\nu=1/5$ are easily replaced by charge-density-wave (CDW) or Wigner-crystal order~\cite{PhysRevLett.108.126405,PhysRevB.86.205125,PhysRevLett.111.126802,PhysRevB.104.085107}. At lower density, repulsive interactions favor spatial modulation once kinetic energy is quenched. Hence, $\nu=1/5$ sensitively tests a Chern band\'s ability to support a dilute fractional liquid. Flatness and a nonzero Chern number are insufficient; the band-projected interaction must also resemble its lowest-Landau-level counterpart. The Berry curvature $\Omega(\boldsymbol{k})$ and Fubini--Study metric $g(\boldsymbol{k})$ characterize this quantum geometry~\cite{PhysRevLett.127.246403,LIN2026100339,PhysRevB.104.045103,PhysRevX.14.041040,PhysRevLett.93.206602,RevModPhys.82.1539,PhysRevB.81.245129,Ma_2013}. In the ideal limit, $\mathrm{Tr}\,g(\boldsymbol{k})=|\Omega(\boldsymbol{k})|$ throughout the Brillouin zone. Departures from this trace condition modify the projected-interaction form factors and can favor competing symmetry-broken states~\cite{PhysRevB.90.165139,PhysRevB.103.125406,PhysRevB.104.085107,PhysRevResearch.5.L012015}.
	The question is whether quantum geometry is sufficiently Landau-level-like to support a stable dilute fractional liquid.
	
	In this Letter, we use a center-decorated kagome lattice, in which the center-site hopping $t_2$ independently tunes the quantum geometry of an isolated $C=1$ flat band~\cite{PhysRevB.94.081102,PhysRevLett.110.106804}.
	Tuning $t_2$ redistributes the band wave functions, improves the trace condition, and makes the projected-interaction form factors more Landau-level-like, correlating with enhanced FCI stability in the regime where charge ordering is a natural competing tendency.
	Exact diagonalization identifies a geometry-improved window that substantially stabilizes Laughlin-like states at both $\nu=1/3$ and $\nu=1/5$.
	
\begin{figure}[t!]
		\centering
		\includegraphics[width=0.46\textwidth,height=0.40\textheight,keepaspectratio]{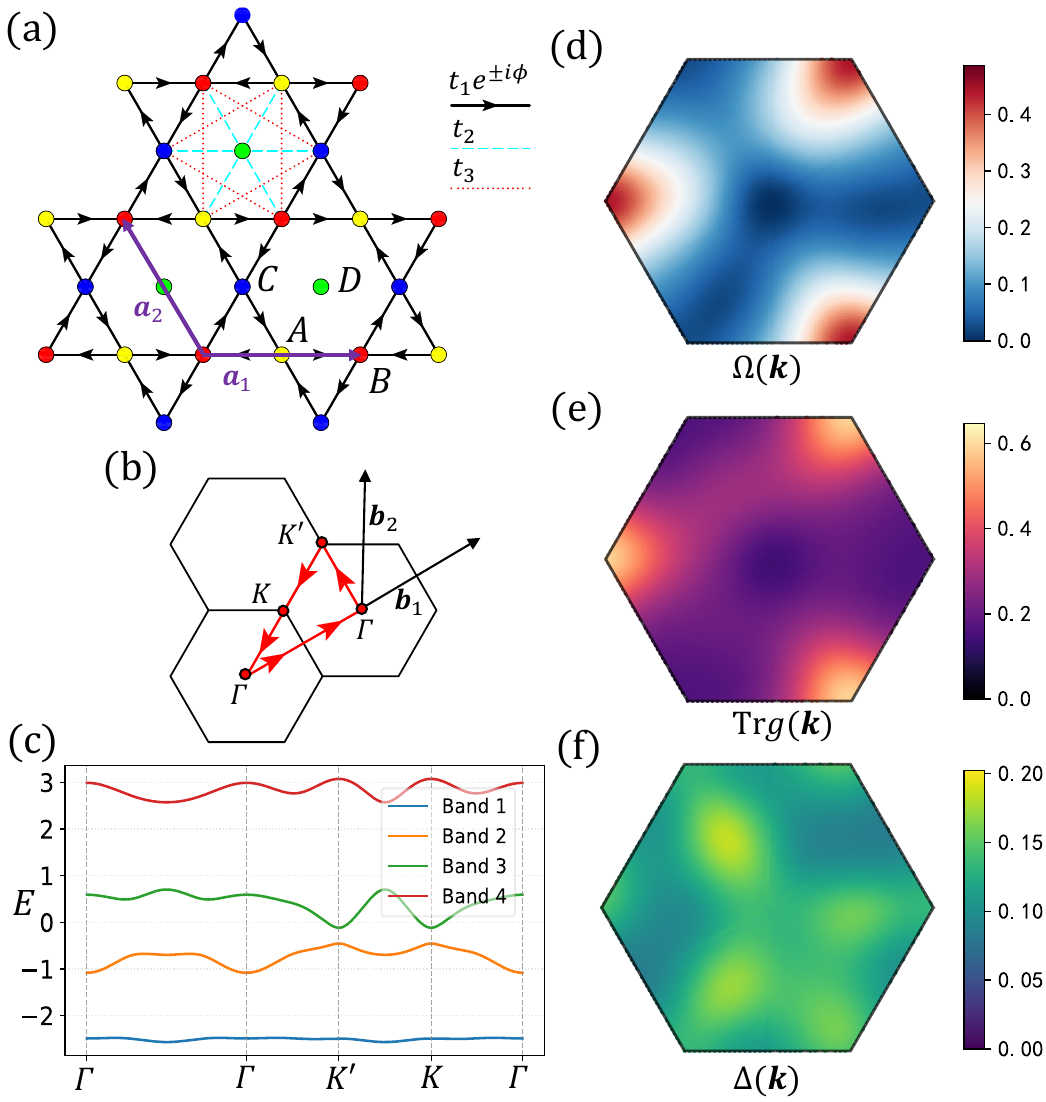}
		\caption{
			Center-decorated kagome lattice and single-particle band geometry. (a) Lattice structure and hopping pattern. (b) Brillouin zone. (c) Representative flat lowest band with Chern number $C=1$ along the high-symmetry path, evaluated at $t_1=1$, $t_2=-1/3$, $t_3=1/3$, and $\phi\approx0.8\pi$. (d)-(f) Berry curvature, trace of the quantum metric, and trace-condition deviation $\Delta(\boldsymbol{k})=\mathrm{Tr}\,g(\boldsymbol{k})-|\Omega(\boldsymbol{k})|$.
		}
		\label{fig:model}
	\end{figure}

	\parhead{Decorated-kagome band geometry} The lattice contains kagome sublattices $A$, $B$, and $C$ plus a center site $D$ in each hexagon [Fig.~\ref{fig:model}(a)]. It can be realized as an ultracold-fermion optical superlattice formed from two commensurate triangular lattices. The center site is essential rather than a passive addition, its hybridization with the kagome backbone, controlled by $t_2$, redistributes the lowest-band wave functions among the four sublattices. It consequently changes the Berry curvature, quantum metric, and projected interaction form factors without changing the lattice topology.
	The Hamiltonian contains the noninteracting part and the interaction part,
	\begin{equation}
		\begin{aligned}
			H_0 &= \sum_{n,m}\sum_{\alpha,\beta} t_{nm}^{\alpha\beta} c^\dagger_{n\alpha} c_{m\beta} + \mathrm{H.c.},\\
			H_{\mathrm{int}} &= U_1 \sum_{\langle n\alpha,m\beta\rangle} n_{n\alpha} n_{m\beta}
			+ U_2 \sum_{\langle\!\langle\!\langle n\alpha,m\beta\rangle\!\rangle\!\rangle} n_{n\alpha} n_{m\beta}.
		\end{aligned}
		\label{eq:H}
	\end{equation}
	Here $c^\dagger_{n\alpha}$ creates a spinless fermion on sublattice $\alpha$ in unit cell $n$, and $n_{n\alpha}=c^\dagger_{n\alpha}c_{n\alpha}$. We first discuss the noninteracting part $H_0$. The nearest-neighbor hoppings on the kagome backbone carry phases $\pm\phi$, the hopping between the center site and the outer sites is $t_2$, and the additional outer-sublattice hopping is $t_3$.
	
	The role of the added $D$ site is not simply to increase the number of bands~\cite{PhysRevB.94.081102,PhysRevLett.110.106804}. It hybridizes with the kagome backbone and changes the distribution of the lowest-band wave functions among the four sublattices. As a result, varying $t_2$ modifies the band form factors that enter the projected interaction, and hence changes the Berry curvature and quantum metric without requiring a different lattice topology.
	The parameter $t_3$ and phase $\phi$ are introduced to break time-reversal symmetry~\cite{PhysRevLett.61.2015}, yielding an isolated flat band with Chern number $C=1$, while $t_2$ is then left as the principal knob for moving within that regime. This separation makes it possible to distinguish flatness optimization from geometry optimization.
	
	After Fourier transformation, the noninteracting Hamiltonian becomes a four-band Bloch matrix $\mathscr{H}(\boldsymbol{k})$, whose explicit off-diagonal elements are given in Supplemental Material Sec.~S1~\cite{supp}. The parameter $t_2$ continuously connects the decorated model to the kagome limit and serves as our main control knob for the band geometry.

	Fixing $t_1=1$, we search the noninteracting parameter space and use the flatness ratio $f=\Delta_{\mathrm{gap}}/W$, namely the gap above the lowest band divided by the bandwidth of that band, as the primary criterion for selecting a reference point. For parameters near $\phi\approx0.8\pi$ and $t_3=1/3$, this procedure yields a representative value $t_2\approx-1/3$. Because the numerical scan is discrete, this choice should be regarded as a convenient near-optimal flat-band point rather than an exact mathematical optimum. The lowest band at this point is nearly flat and topological with $C=1$ [Fig.~\ref{fig:model}(c)].
	
\begin{figure}[t!]
		\centering
		\includegraphics[width=0.46\textwidth,height=0.40\textheight,keepaspectratio]{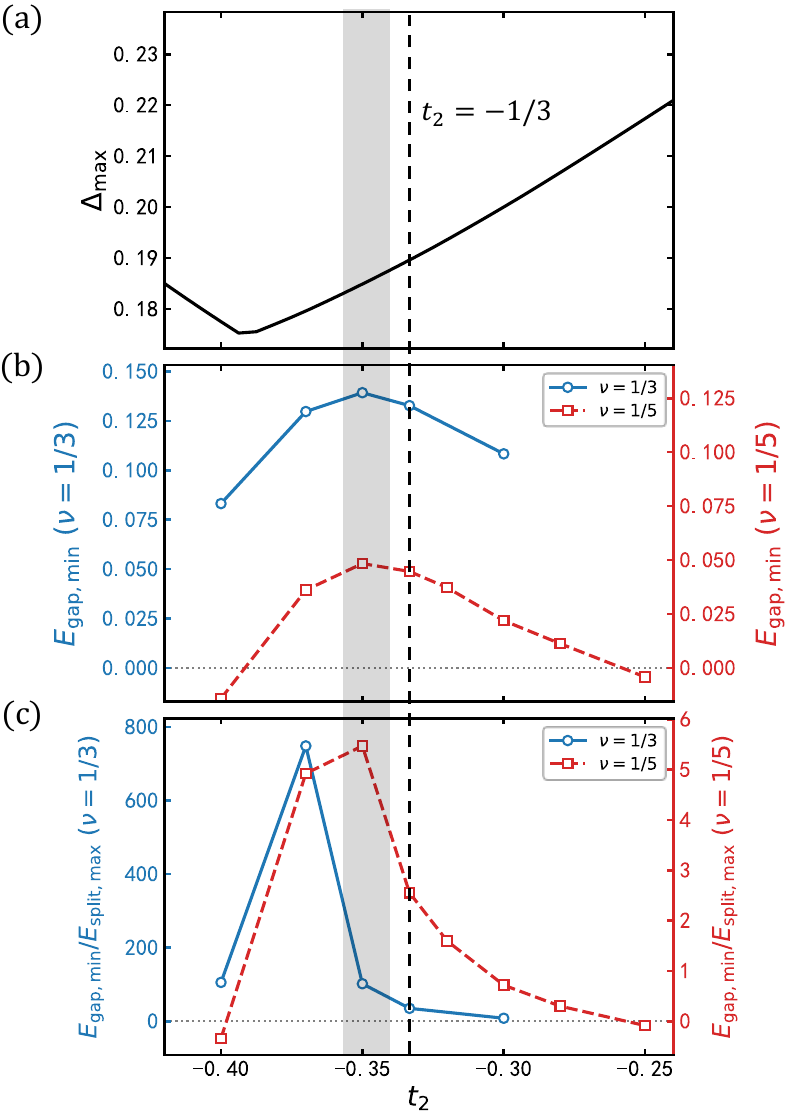}
		\caption{
			Optimization with respect to the center-site hopping $t_2$ between the added $D$ site and the kagome backbone.
			(a) Maximum trace-condition deviation $\Delta_{\max}$. (b) Minimum many-body gap and (c) gap-to-splitting ratio over the twist-angle torus for $\nu=1/3$ and $\nu=1/5$. The shaded region marks the stability-enhanced, geometry-improved window, and the vertical black dashed line indicates $t_2=-1/3$, $t_2=-0.35$ is used below as a representative point.
		}
		\label{fig:gap-delta}
	\end{figure}

	Starting from this flat-band reference point, we then vary $t_2$ continuously to tune the quantum geometry. The corresponding $\Omega(\boldsymbol{k})$ and $\mathrm{Tr}\,g(\boldsymbol{k})$ remain relatively uniform, and the deviation $\Delta(\boldsymbol{k})=\mathrm{Tr}\,g(\boldsymbol{k})-|\Omega(\boldsymbol{k})|$ stays modest across the Brillouin zone [Figs.~\ref{fig:model}(d)-\ref{fig:model}(f)]. More importantly, the scan in Fig.~\ref{fig:gap-delta} shows that the geometry is further improved around $t_2\approx-0.35$, where the lowest band approaches the ideal trace condition more closely than at the flatness-selected starting point.
	This distinction shows that the flatness ratio and the quantum geometry need not be optimized at the same parameter point~\cite{WOS:000366300900001,PhysRevB.90.165139,PhysRevB.104.045103}. In the present model, the center-site hybridization allows the band dispersion and the quantum metric to respond differently as $t_2$ is varied. The many-body optimum can therefore occur slightly away from the flatness reference point, providing a concrete example in which geometry optimization is not merely a refinement of bandwidth minimization.
	
	\parhead{Geometry optimization} To quantify the relation between band geometry and many-body stability, we scan $t_2$ and compare $\Delta_{\max}=\max_{\boldsymbol{k}} \Delta(\boldsymbol{k})$ with many-body quantities extracted under twisted boundary conditions. Figure~\ref{fig:gap-delta}(a) shows that $\Delta_{\max}$ develops a pronounced minimum near $t_2\approx -0.35$. Figures~\ref{fig:gap-delta}(b) and \ref{fig:gap-delta}(c) present, for both $\nu=1/3$ and $\nu=1/5$, the minimum many-body gap $E_{\mathrm{gap},{\min}}$ and the ratio $E_{\mathrm{gap},{\min}}/E_{\mathrm{split},{\max}}$ over the full $(\theta_1,\theta_2)\in[0,2\pi]\times[0,2\pi]$ torus. Here $E_{\mathrm{split},{\max}}$ denotes the maximum splitting within the quasi-degenerate ground-state manifold, and the extrema of the gap and splitting need not occur at the same twist angle. The flux torus is discretized with $11\times11$ points for $\nu=1/3$ and $16\times16$ points for $\nu=1/5$,
	respectively. For both fillings, moving away from the flatness-selected reference point toward the geometry-improved window around $t_2\approx -0.35$ substantially enhances the many-body stability. The strongest many-body response does not need to occur exactly at the point of minimal geometric deviation, but the overall enhancement remains concentrated in the same geometry-improved region, providing strong evidence that improved band geometry promotes FCI stability.
	
	We compare representative points within the same geometry-improved window in Supplemental Material Sec.~S2~\cite{supp}, and we show the spectral flow at the flat-band reference point in Sec.~S3. We use $t_2=-0.35$ below as a representative optimized parameter because it captures the enhanced gap, weak ground-state splitting, smooth momentum occupation, and suppressed finite-momentum structure factor features in a single point.
	
	\parhead{Many-body spectra and flow} We now return to the interacting problem. Since the gap above the lowest single-particle band is much larger than both the bandwidth and the interaction scale in the parameter regime of interest, we project the full Hamiltonian into the lowest $C=1$ band~\cite{PhysRevX.1.021014,PhysRevLett.106.236804,WOS:000294805300016,Liu_2024}. The projected many-body Hamiltonian can be written as
	\begin{equation}
		\begin{aligned}
			H_p =& \sum_{\boldsymbol{k}\in \mathrm{1BZ}} \epsilon_{\boldsymbol{k}} \gamma^\dagger_{\boldsymbol{k}}\gamma_{\boldsymbol{k}} \\
			&+ \sum_{\{\boldsymbol{k}_i\}\in \mathrm{1BZ}} V(\boldsymbol{k}_1,\boldsymbol{k}_2,\boldsymbol{k}_3,\boldsymbol{k}_4)
			\gamma^\dagger_{\boldsymbol{k}_1}\gamma^\dagger_{\boldsymbol{k}_2}\gamma_{\boldsymbol{k}_3}\gamma_{\boldsymbol{k}_4},
		\end{aligned}
		\label{eq:Hp}
	\end{equation}
	where the interaction matrix elements inherit the geometric form factors of the lowest band. We solve this problem by exact diagonalization on finite tori with $N_u=N_1\times N_2$ unit cells. Following the physical trend that shorter-ranged repulsion favors the denser state while longer-ranged repulsion is more relevant at lower filling, we use $U_1=1$, $U_2=0$ for $\nu=1/3$, and $U_1=0$, $U_2=2$ for $\nu=1/5$ as representative interaction profiles. For $\nu=1/5$, the stability-enhanced window around $t_2\approx -0.35$ remains robust for nearby interaction choices, including both changes in the dominant third-neighbor repulsion and weak admixtures of nearest-neighbor repulsion, as documented in Supplemental Material Secs.~S4 and S5~\cite{supp}.
	
	On a torus, an Abelian Laughlin state at filling $\nu=1/q$ is expected to form a $q$-fold quasi-degenerate ground-state manifold~\cite{PhysRevX.1.021014,PhysRevB.85.075128}. The finite-size splitting inside this manifold is not itself a sign of symmetry breaking, provided that it remains small compared with the gap to higher excitations and decreases or stays controlled across accessible clusters. Flux insertion gives a complementary test, the individual low-energy states need not return to themselves after one flux quantum, but the full manifold should be permuted internally and remain separated from excited states~\cite{PhysRevX.1.021014,PhysRevB.85.075128}. These two diagnostics are especially useful at $\nu=1/5$, where a competing charge-ordered phase can also produce low-lying states but would not show the same topological spectral flow.
	
\begin{figure}[!htbp]
		\centering
		\includegraphics[width=0.46\textwidth,height=0.40\textheight,keepaspectratio]{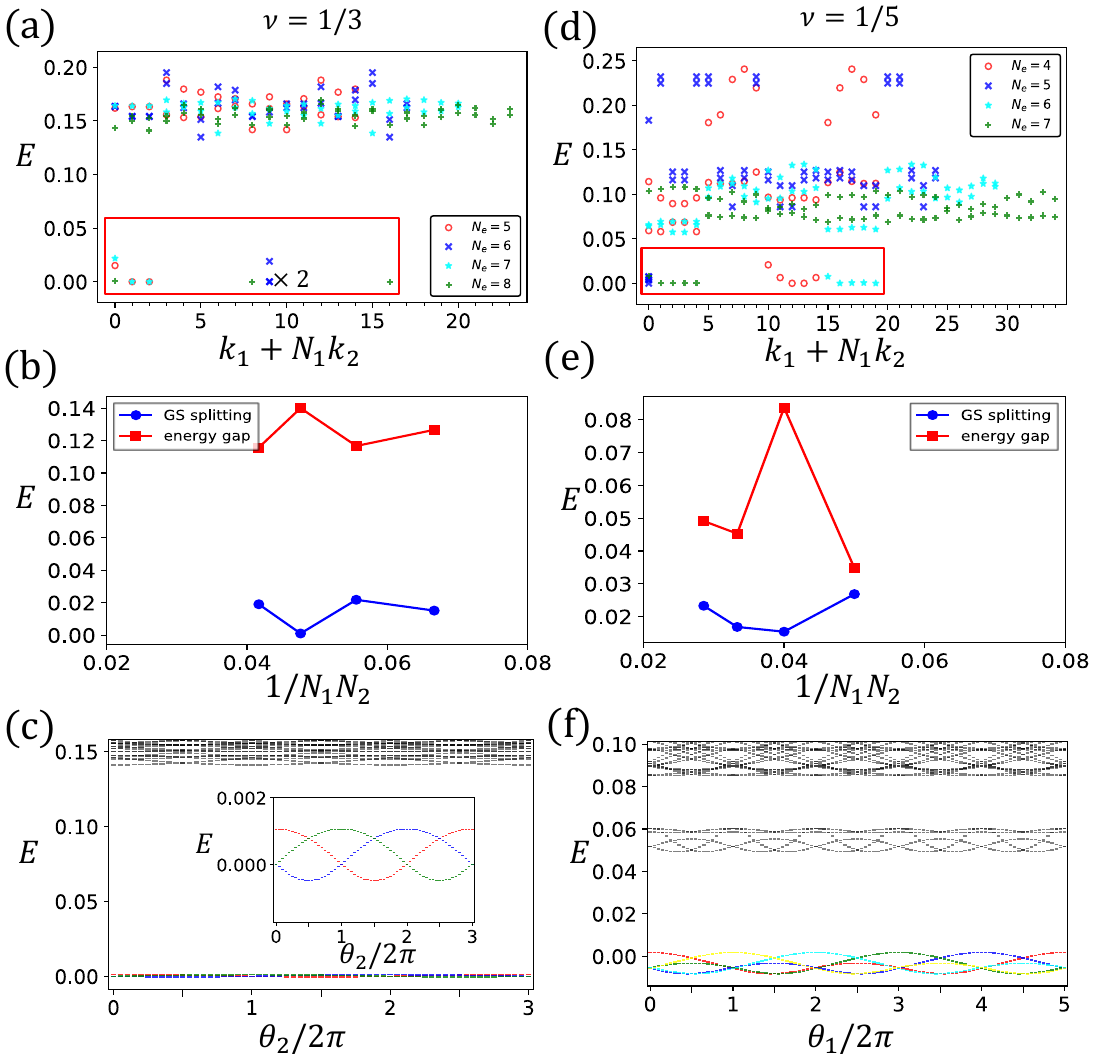}
		\caption{
			Low-energy spectra, finite-size scaling, and spectral flow at the optimized parameter. Panels (a)--(c) show results for $\nu=1/3$, and panels (d)--(f) show results for $\nu=1/5$: (a),(d) low-energy spectra, (b),(e) finite-size trends of the ground-state splitting and excitation gap, and (c),(f) spectral flow under flux insertion. The quasi-degenerate ground-state manifolds remain separated from higher excitations and return to themselves under flux insertion.
		}
		\label{fig:size-spectrumflow}
	\end{figure}

	Figure~\ref{fig:size-spectrumflow} shows the corresponding low-energy spectra, finite-size trends, and spectral flow. The expected threefold and fivefold low-energy manifolds appear clearly at $\nu=1/3$ and $\nu=1/5$, respectively, and are highlighted by red boxes in the spectra. Their momentum-sector distributions are consistent with the generalized Pauli principle for Abelian Laughlin states~\cite{PhysRevLett.100.246802,PhysRevX.1.021014,PhysRevB.85.075128}. The splitting within each manifold stays small while a finite gap to higher excited states survives across the accessible system sizes. Under flux insertion, the ground-state manifold evolves without mixing with higher states and returns to itself after the insertion of three flux quanta for $\nu=1/3$ and five flux quanta for $\nu=1/5$, consistent with Abelian Laughlin-like FCIs.
	
\begin{figure}[t!]
		\centering
		\includegraphics[width=0.46\textwidth,height=0.40\textheight,keepaspectratio]{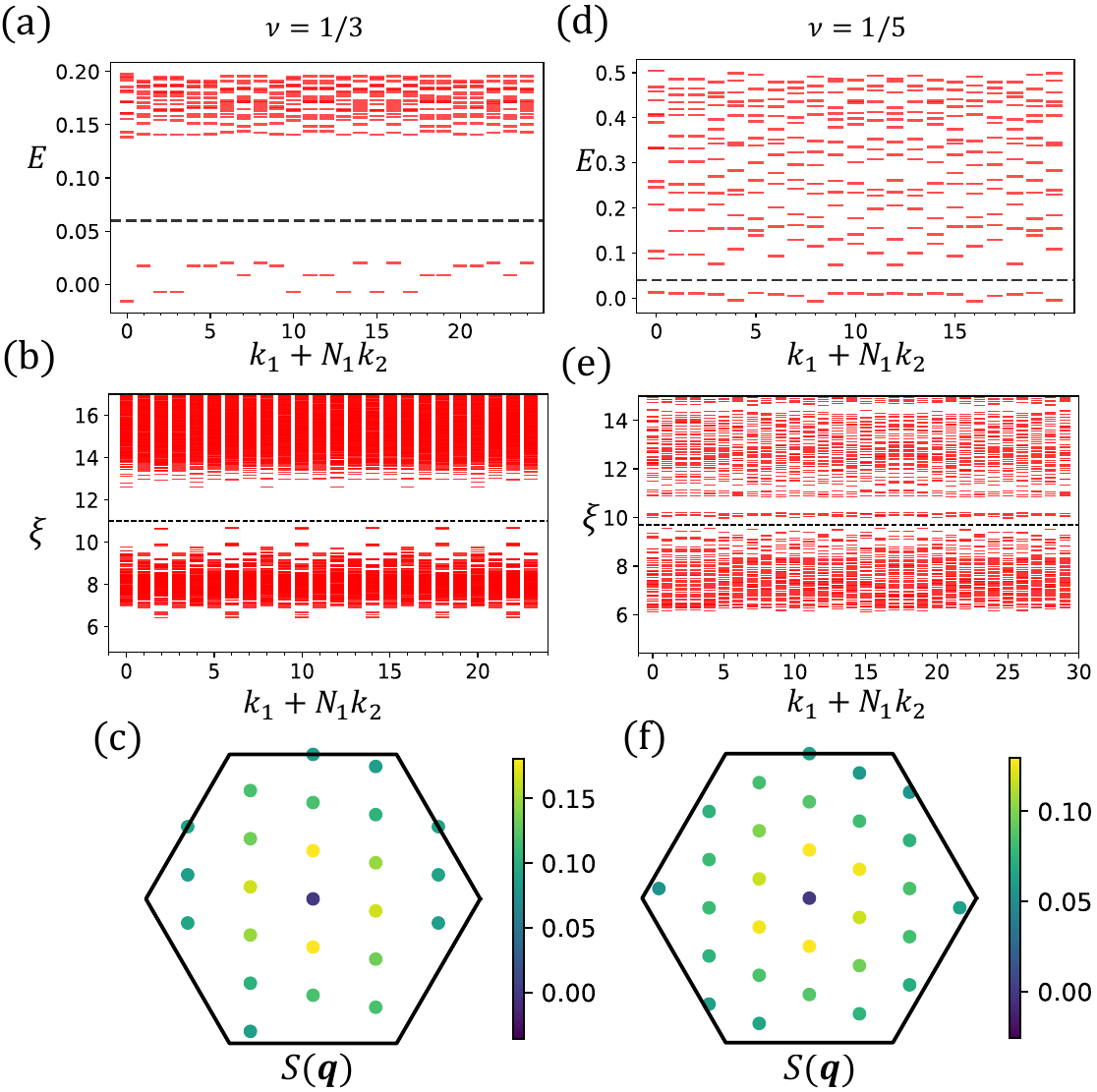}
		\caption{
			Quasihole spectra, particle entanglement spectra, and static structure factors at the optimized parameter. Panels (a)--(c) correspond to $\nu=1/3$, and panels (d)--(f) correspond to $\nu=1/5$: (a),(d) quasihole spectra, (b),(e) particle entanglement spectra, and (c),(f) static structure factors $S(\boldsymbol{q})$. The counting agrees with Abelian Laughlin FCIs, while $S(\boldsymbol{q})$ shows no sharp ordering peak.
		}
		\label{fig:ex-pes-sq}
	\end{figure}

	\parhead{Topological characterization} Further evidence for topological order is collected in Fig.~\ref{fig:ex-pes-sq}. The quasihole excitation spectra [Figs.~\ref{fig:ex-pes-sq}(a) and \ref{fig:ex-pes-sq}(d)] exhibit clear low-energy manifolds separated by a gap, and the number of states below the gap matches the counting rules of the generalized Pauli principle for Abelian Laughlin states~\cite{PhysRevLett.100.246802,PhysRevX.1.021014,PhysRevB.85.075128}. The particle entanglement spectra (PES) [Figs.~\ref{fig:ex-pes-sq}(b) and \ref{fig:ex-pes-sq}(e)] show robust entanglement gaps, with the correct low-lying counting $2730$ for $\nu=1/3$ and $1360$ for $\nu=1/5$. Finally, the static structure factors $S(\boldsymbol{q})$ [Figs.~\ref{fig:ex-pes-sq}(c) and \ref{fig:ex-pes-sq}(f)] display no pronounced Bragg peaks, indicating no clear tendency toward CDW or Wigner-crystal order in the optimized parameter regime.
	The three diagnostics probe different possible failure modes. Quasihole counting tests whether removing particles or adding flux produces the characteristic exclusion statistics of the Laughlin sequence. The PES is more stringent because it probes the internal structure of the ground-state manifold rather than only the energy ordering; the observed low-lying counting indicates that the optimized states retain the expected topological entanglement fingerprint. The structure factor $S(\boldsymbol{q})$, by contrast, is designed to rule out the most natural competing interpretation at low filling. The absence of sharp finite-momentum peaks shows that the enhanced stability near $t_2=-0.35$ is not obtained by forming a simple density-ordered state.
	
	\parhead{Additional diagnostics} The momentum-space occupation in Supplemental Material Sec.~S1~\cite{supp} fluctuates only weakly around the average filling for both $\nu=1/3$ and $\nu=1/5$, consistent with a spatially uniform state rather than a CDW phase. We also evaluate the many-body Chern number from the Berry curvature on the twisted-boundary-condition torus~\cite{PhysRevB.31.3372,PhysRevB.48.8890,PhysRevLett.133.066601},
	\begin{equation}
		C = \frac{1}{2\pi}\int \mathrm{d}\theta_1 \mathrm{d}\theta_2\, F(\theta_1,\theta_2),
		\label{eq:Chern}
	\end{equation}
	and find an exactly quantized total many-body Chern number $C_{\mathrm{tot}}=1$ for the nearly degenerate ground-state manifold at both fillings, in perfect agreement with the expected Laughlin-like FCI response. For $\nu=1/5$, the enhancement near the geometry-improved window survives across nearby interaction profiles, as further shown in Supplemental Material Secs.~S4 and S5~\cite{supp}.
	The Chern-number calculation is performed for the entire nearly degenerate manifold, rather than for a single finite-size eigenstate, because the individual states are mixed by flux insertion on the torus. The quantized total value therefore provides a global response diagnostic that complements the local spectral and counting information. Taken together, the gap, spectral flow, quasihole counting, entanglement counting, smooth occupation, absence of Bragg peaks, and total Chern number give a mutually consistent picture, the optimized region supports a fractional topological liquid rather than a finite-size remnant of charge order.
	
	\parhead{Experimental implementation} The proposed mechanism can be tested in an ultracold-fermion optical superlattice. Band mapping or lattice-modulation spectroscopy can characterize the single-particle geometry, incompressibility can probe the many-body gap, quantized center-of-mass displacement can reveal the Hall response, and Bragg spectroscopy of $S(\boldsymbol{q})$ can test the absence of charge order~\cite{PhysRevLett.125.236401,Nature.619.495}. A concrete route is a superlattice built from two commensurate triangular lattices, whose relative displacement and depths generate a four-site unit cell with three kagome sites $A,B,C$ and one hexagon-center site $D$~\cite{PhysRevLett.108.045305}. Fermionic alkaline-earth-like atoms such as $^{87}\text{Sr}$ or $^{173}\text{Yb}$ are natural candidates; a single nuclear-spin component of the electronic ground state $^1S_0$ realizes the spinless fermions used in the model, while the long-lived clock state $^3P_0$ may be used for state-dependent addressing or Raman coupling when needed~\cite{PhysRevLett.101.170504,PhysRevLett.120.143601}. Alkali fermions such as $^{40}\text{K}$ or $^{6}\text{Li}$ provide an alternative implementation with Raman-assisted tunneling and tunable interactions~\cite{PhysRevLett.94.086803,PhysRevA.76.023613,PhysRevLett.126.103201}.

	In such a superlattice the relevant laser parameters are most naturally specified in recoil-energy units, with $E_R=\hbar^2 k_L^2/(2m)$ set by the lattice wave vector $k_L$ and atomic mass $m$. Typical shallow-to-intermediate lattice depths $V_{\rm s},V_{\rm l}\sim 3$--$10E_R$ keep the lowest band well separated while retaining appreciable tunneling. The relative superlattice phase fixes the center-site offset $\Delta_D$, and tuning $\Delta_D/E_R$ together with the long-wavelength lattice depth controls the hybridization between $D$ and the kagome backbone. After calibration by lattice modulation or band spectroscopy, this realizes the effective hopping ratios used here, with $t_1$ setting the energy unit and $t_2/t_1$ tuned through the geometry-improved window near $-0.35$. The additional hopping $t_3/t_1\simeq 1/3$ and the Peierls phase $\phi\simeq0.8\pi$ can be generated by Raman-assisted tunneling, lattice shaking, spin-dependent potentials, or related Floquet schemes, with the modulation or Raman detuning chosen large compared with the tunneling and interaction scales but small enough to avoid unwanted interband excitation~\cite{WOS:000344631400043,WOS:000494944200009,PhysRevLett.126.103201,PhysRevA.111.043315,PhysRevA.87.023622}. The superlattice geometry and hopping parameters can be calibrated by Kapitza--Dirac diffraction, lattice modulation spectroscopy, and time-of-flight band mapping.

	\parhead{Conclusion} We have shown that a center-decorated kagome lattice can stabilize dilute fractional Chern insulators by tuning the quantum geometry of an isolated $C=1$ band.
	Moving from the flatness-selected reference point near $t_2=-1/3$ into the geometry-improved window around $t_2\approx-0.35$ reduces the trace-condition deviation, bringing the projected interaction form factors closer to the Landau-level limit and enhancing the stability of the fractional topological liquid.
	Stabilization here means an enhanced excitation gap relative to the ground-state splitting, together with a topological ground-state manifold that remains isolated under flux insertion. The spectra, flux response, counting diagnostics, structure factors, and total many-body Chern number consistently identify Laughlin-like fractional Chern insulators at both $\nu=1/3$ and $\nu=1/5$, while the persistence of the $\nu=1/5$ enhancement under nearby interaction profiles shows that the effect is tied to a geometry-improved regime rather than to a single fine-tuned point.

	Our results suggest a practical design principle for dilute fractional Chern insulators, instead of optimizing band flatness alone, one should also engineer band geometries that more closely satisfy the trace condition. The center-decorated kagome construction provides a compact route to this goal because the additional site degree of freedom tunes the band geometry while preserving a simple lattice connectivity. This is particularly important for dilute fractional Chern insulators, whose competition with charge-ordered phases makes them a sensitive test of whether a Chern band has the geometric structure needed to support a fractional liquid. This principle can be explored in programmable optical lattices and, more broadly, in decorated Chern-band platforms.
	
	\begin{acknowledgments}
		Y.X. Ding, W.T. Li, and W.M. Liu are supported by the National Key R\&D Program of China under grants Nos. 2024YFF0726700, 2021YFA1400900, 2021YFA0718300, NSFC under grants Nos. 12334012, 12234012, 52327808, Space Application System of China Manned Space Program, Elite Revitalizing Inner Mongolia Program (2025TGL05).
	\end{acknowledgments}
	
	\bibliographystyle{apsrev4-2}
	\bibliography{references,sm/smreferences}
	
	\clearpage
	\onecolumngrid 
	
	\setcounter{equation}{0}
	\setcounter{figure}{0}
	\setcounter{table}{0}
	\setcounter{page}{1}
	\setcounter{section}{0}
	\setcounter{secnumdepth}{2} 
	
	\makeatletter
	\renewcommand{\theequation}{S\arabic{equation}}
	\renewcommand{\thefigure}{S\arabic{figure}}
	\renewcommand{\thetable}{S\arabic{table}}
	\renewcommand{\thesection}{S\arabic{section}}
	\renewcommand{\thepage}{S\arabic{page}}
	\makeatother
	
	\begin{center}
		{\large\bfseries Supplemental Material: ``Quantum-geometry stabilization of dilute fractional Chern insulators''\par}
		\vspace{1.0em}
		{Ying-Xing Ding,$^{1,2}$ Li-Min Zhang,$^{3}$ Wen-Tong Li,$^{1,2}$ D. L. Zhou,$^{1,2}$ and Wu-Ming Liu$^{1,*}$\par}
		\vspace{0.8em}
		{\small $^{1}$\textit{Beijing National Laboratory for Condensed Matter Physics, Institute of Physics, Chinese Academy of Sciences, Beijing 100190, China}\par}
		{\small $^{2}$\textit{School of Physical Sciences, University of Chinese Academy of Sciences, Beijing 100049, China}\par}
		{\small $^{3}$\textit{CAS Key Laboratory of Quantum Information, University of Science and Technology of China, Hefei 230026, China}\par}
		\vspace{0.5em}
		{\small $^*$Electronic address: wliu@iphy.ac.cn\par}
	\end{center}
	\vspace{1.5em}
	

In this Supplemental Material, we provide additional numerical diagnostics for the geometry-optimized center-decorated kagome model. Section~\ref{sec:nk-berry} presents the momentum-space occupation and plaquette Berry phase at $t_2=-0.35$, supporting the interpretation of the $\nu=1/3$ and $\nu=1/5$ states as fractional Chern insulators rather than charge-ordered phases~\cite{PhysRevLett.106.236802,PhysRevLett.106.236803,PhysRevLett.106.236804,WOS:000294805300016,PhysRevB.103.125406,PhysRevB.104.085107}. Section~\ref{sec:compare} compares the nearby parameters $t_2=-0.35$ and $t_2=-0.37$ in terms of momentum occupation and static structure factor. Section~\ref{sec:spectral-flow} shows the spectral flow at the flat-band reference point $t_2=-1/3$, illustrating how the geometry-optimized window improves the many-body response~\cite{PhysRevLett.127.246403,PhysRevB.104.045103}. Sections~\ref{sec:u2-robust} and \ref{sec:u1-robust} examine the robustness of the fragile $\nu=1/5$ state against nearby interaction profiles. We show that the stability-enhanced window near the geometry-improved region around $t_2\approx -0.35$ persists when the dominant third-neighbor repulsion is varied and when weak nearest-neighbor repulsion is admixed.

\section{Additional momentum-space diagnostics and many-body Berry phase}\label{sec:nk-berry}

For completeness, we first give the explicit Bloch Hamiltonian used in the main text. After Fourier transformation, the noninteracting Hamiltonian takes the form $\mathscr{H}(\boldsymbol{k})$. Letting sublattices $A$, $B$, $C$, and $D$ correspond to indices $1$, $2$, $3$, and $4$, the independent off-diagonal matrix elements are
\begin{subequations}\label{eq:Hk-sm}
	\begin{align}
		\mathscr{H}_{21}(\boldsymbol{k})=&t_1 e^{-i\phi}(1+e^{ik_1})+t_3(e^{-ik_2}+e^{i(k_1+k_2)}),\\
		\mathscr{H}_{31}(\boldsymbol{k})=&t_1 e^{i\phi}(1+e^{i(k_1+k_2)})+t_3(e^{ik_1}+e^{ik_2}),\\
		\mathscr{H}_{32}(\boldsymbol{k})=&t_1 e^{-i\phi}(1+e^{ik_2})+t_3(e^{i(k_1+k_2)}+e^{-ik_1}),\\
		\mathscr{H}_{41}(\boldsymbol{k})=&t_2 (e^{i(k_1+k_2)}+e^{ik_1}),\\
		\mathscr{H}_{42}(\boldsymbol{k})=&t_2 (1+e^{i(k_1+k_2)}),\\
		\mathscr{H}_{43}(\boldsymbol{k})=&t_2 (1+e^{ik_1}),
	\end{align}
\end{subequations}
with $k_i=\boldsymbol{k}\cdot\boldsymbol{a}_i$ and $\mathscr{H}_{ij}(\boldsymbol{k})=[\mathscr{H}_{ji}(\boldsymbol{k})]^*$.

\begin{figure}[htbp]
	\centering
	\includegraphics[width=0.8\textwidth]{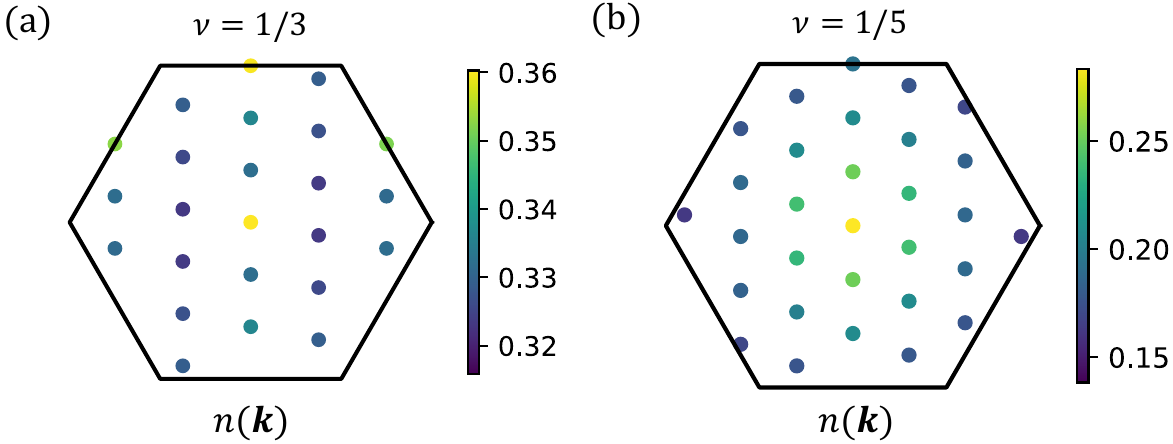}
	\caption{Momentum-space occupation $n(\boldsymbol{k})$ for the optimized parameter $t_2=-0.35$. The two panels correspond to $\nu=1/3$ and $\nu=1/5$, respectively. In both cases, the occupation varies only weakly around the average filling, supporting the interpretation of a uniform FCI state.}
	\label{fig:nk}
\end{figure}

This Supplemental Material collects several diagnostics omitted from the main text. Fig.~\ref{fig:nk} shows the momentum-space occupation $n(\boldsymbol{k})$ for the optimized parameter $t_2=-0.35$ at $\nu=1/3$ and $\nu=1/5$. In both cases, $n(\boldsymbol{k})$ fluctuates only weakly around the average filling, consistent with a spatially uniform liquid rather than a charge-ordered state.

\begin{figure}[htbp]
	\centering
	\includegraphics[width=0.8\textwidth]{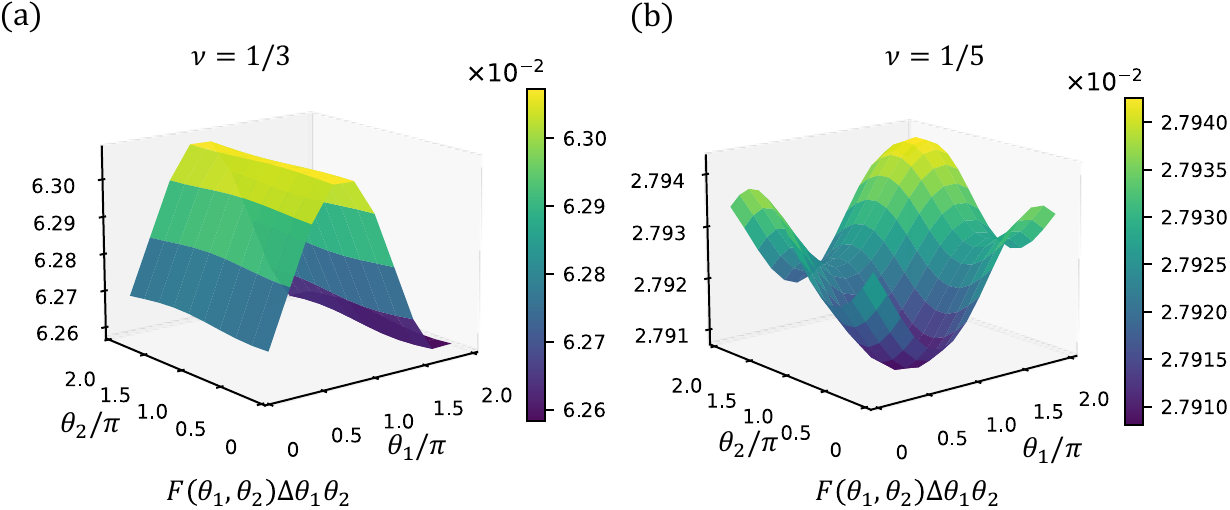}
	\caption{Local many-body Berry phase on the discretized twist-angle torus for the optimized parameter $t_2=-0.35$. The two panels correspond to $\nu=1/3$ and $\nu=1/5$, respectively. The plotted quantity is the Berry phase on each elementary plaquette of the discretized mesh. In both cases, summing over all plaquettes gives the total many-body Chern number $C_{\mathrm{tot}}=1$.}
	\label{fig:berryphase}
\end{figure}

We also evaluate the many-body Berry phase on the discretized twisted-boundary-condition torus~\cite{PhysRevB.31.3372,PhysRevB.48.8890}. The quantity shown in Fig.~\ref{fig:berryphase} is the Berry phase accumulated on each elementary plaquette of the discretized $(\theta_1,\theta_2)$ mesh, rather than the continuous Berry curvature itself. For both $\nu=1/3$ and $\nu=1/5$, the plaquette Berry phase is distributed smoothly over the flux torus, without pronounced singular features. Summing over the full mesh yields the exactly quantized total many-body Chern number $C_{\mathrm{tot}}=1$ for the nearly degenerate ground-state manifold~\cite{PhysRevX.1.021014}.

\FloatBarrier

\section{Comparison between nearby parameters at one-third filling}\label{sec:compare}

\begin{figure}[htbp]
	\centering
	\includegraphics[width=0.82\textwidth]{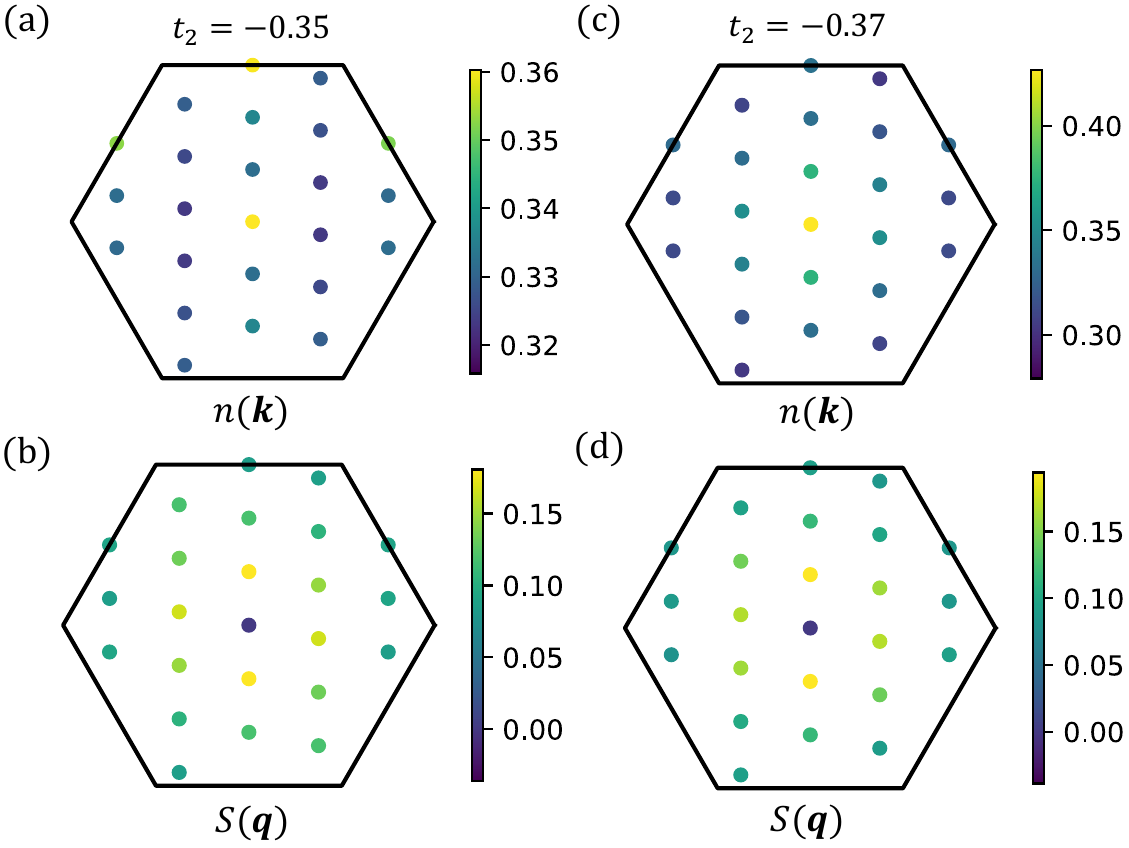}
	\caption{Direct comparison between the nearby parameters $t_2=-0.35$ and $t_2=-0.37$ at $\nu=1/3$. Panels (a) and (b) show the momentum-space occupation $n(\boldsymbol{k})$ and the static structure factor $S(\boldsymbol{q})$ at $t_2=-0.35$, respectively, while panels (c) and (d) show the corresponding quantities at $t_2=-0.37$. Although $t_2\approx -0.37$ gives a slightly sharper optimum in the ratio $\mathrm{gap}_{\min}/\mathrm{split}_{\max}$, the parameter $t_2=-0.35$ exhibits a more uniform momentum occupation and weaker finite-momentum structure-factor features, making it the more balanced representative point.}
	\label{fig:nk-sq-differ}
\end{figure}

As discussed in the main text, the ratio $E_{\mathrm{gap},{\min}}/E_{\mathrm{split},{\max}}$ at $\nu=1/3$ develops a slightly sharper optimum near $t_2\approx -0.37$. Nevertheless, the choice $t_2=-0.35$ provides a better overall compromise once the momentum-space occupation and the static structure factor are considered together. Fig.~\ref{fig:nk-sq-differ} compares the two nearby parameter choices directly. Relative to $t_2=-0.37$, the state at $t_2=-0.35$ shows a more uniform momentum occupation and more strongly suppressed finite-momentum features in $S(\boldsymbol{q})$. This is why we adopt $t_2=-0.35$ as the representative optimized parameter in the main text, consistent with the general sensitivity of dilute FCIs to band geometry~\cite{PhysRevLett.127.246403,PhysRevB.104.045103}.

\section{Spectral flow at the flat-band reference point}\label{sec:spectral-flow}

\begin{figure}[htbp]
	\centering
	\includegraphics[width=0.8\textwidth]{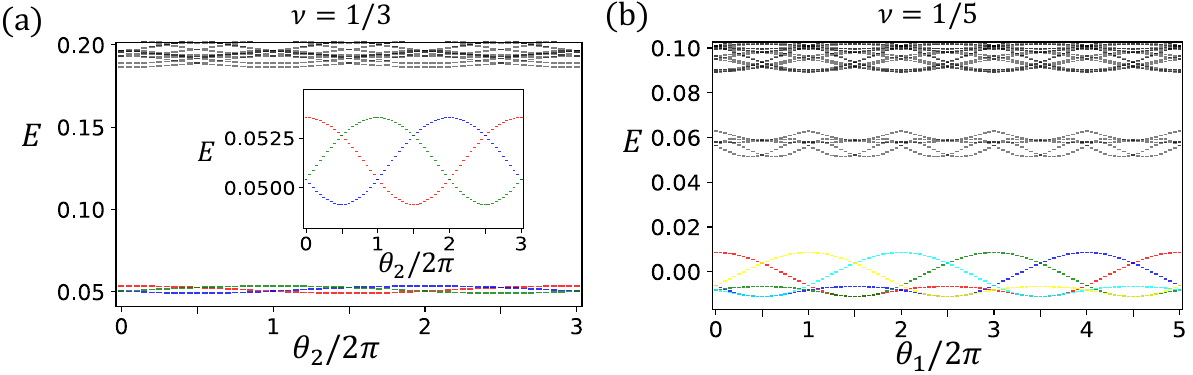}
	\caption{Spectral flow at the flat-band reference point $t_2=-1/3$, before the quantum-geometry optimization toward $t_2=-0.35$. This comparison illustrates that the single-particle flat-band reference point does not coincide with the optimal many-body parameter regime.}
	\label{fig:spect-flow0}
\end{figure}

The main text emphasizes that the flat-band reference point selected from the single-particle flatness ratio near $t_2=-1/3$ is not identical to the optimal many-body parameter. To illustrate this distinction, Fig.~\ref{fig:spect-flow0} presents the spectral flow at the reference point $t_2=-1/3$, before the additional quantum-geometry optimization toward $t_2=-0.35$. Compared with the optimized case shown in the main text, the spectral evolution at $t_2=-1/3$ is less favorable, further supporting the conclusion that the best many-body stability is achieved only after moving away from the flat-band reference point toward the geometry-optimized window.

\clearpage

\section{Robustness against the strength of third-neighbor repulsion}\label{sec:u2-robust}

\begin{figure}[htbp]
	\centering
	\includegraphics[width=0.82\textwidth]{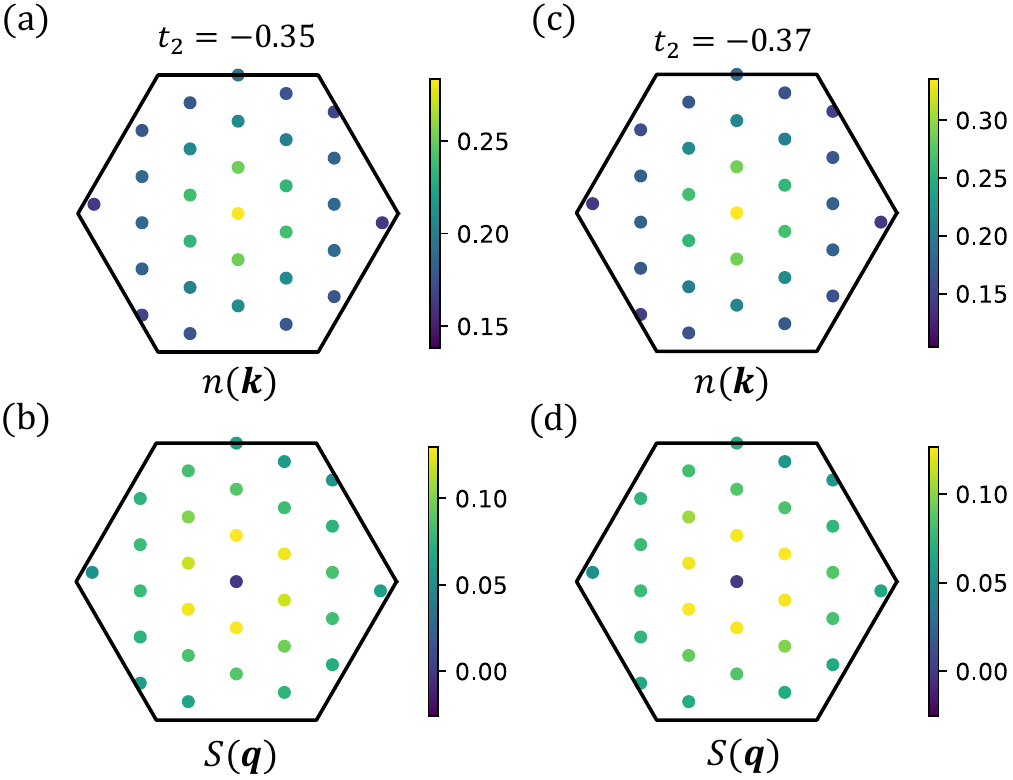}
	\caption{Direct comparison between the nearby parameters $t_2=-0.35$ and $t_2=-0.37$ at $\nu=1/5$ for the representative interaction profile $(U_1,U_2)=(0,2.0)$. Panels (a) and (b) show the momentum-space occupation $n(\boldsymbol{k})$ and the static structure factor $S(\boldsymbol{q})$ at $t_2=-0.35$, respectively, while panels (c) and (d) show the corresponding quantities at $t_2=-0.37$. The structure factors are comparable and show no pronounced ordering peak, while the momentum occupation is smoother at $t_2=-0.35$.}
	\label{fig:nk-sq-differ-fifth}
\end{figure}

We first compare the two nearby geometry-improved parameters for the representative $\nu=1/5$ interaction profile $(U_1,U_2)=(0,2.0)$. Figure~\ref{fig:nk-sq-differ-fifth} shows that the momentum occupation is smoother at $t_2=-0.35$ than at $t_2=-0.37$, while the structure factors at the two nearby points are comparable and show no pronounced ordering peak. This supports using $t_2=-0.35$ as the more balanced representative point of the stability-enhanced, geometry-improved regime.

\begin{figure}[htbp]
	\centering
	\includegraphics[width=0.82\textwidth]{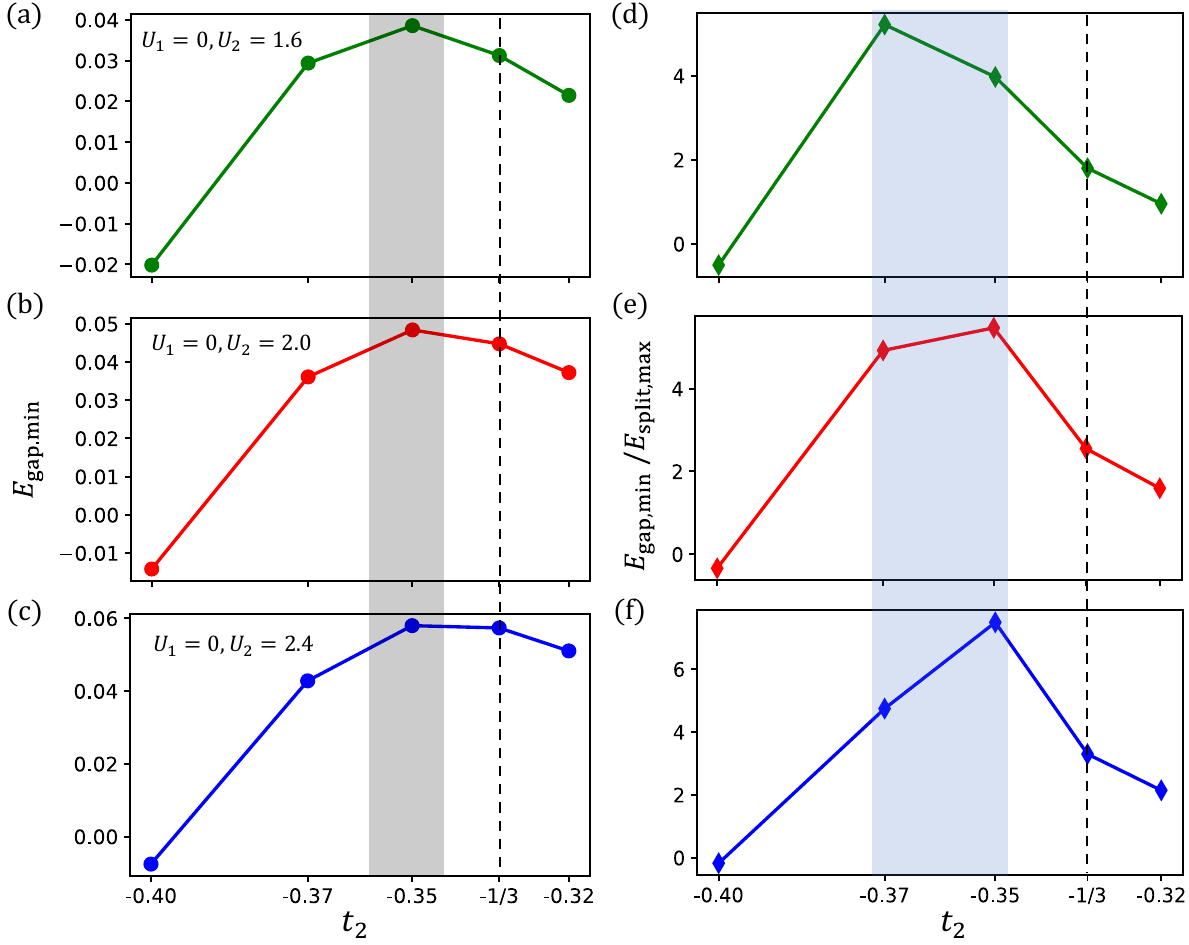}
	\caption{Robustness of the $\nu=1/5$ stability-enhanced window against the strength of the dominant third-neighbor repulsion. Panels (a)--(c) show the minimum many-body gap $E_{\mathrm{gap},\min}$ as a function of $t_2$ for $(U_1,U_2)=(0,1.6)$, $(0,2.0)$, and $(0,2.4)$, respectively. Panels (d)--(f) show the corresponding ratio $E_{\mathrm{gap},\min}/E_{\mathrm{split},\max}$ for the same interaction profiles. The gray marker in the gap panels highlights $t_2=-0.35$, where the minimum gap is maximal. In the ratio panels, the markers indicate the nearby candidate points $t_2=-0.37$ and $t_2=-0.35$.}
	\label{fig:u2-robust}
\end{figure}

We next vary the dominant third-neighbor repulsion while keeping $U_1=0$. Figure~\ref{fig:u2-robust} compares $(U_1,U_2)=(0,1.6)$, $(0,2.0)$, and $(0,2.4)$. For each interaction profile, we compute the minimum many-body gap $E_{\mathrm{gap},\min}$ and the ratio $E_{\mathrm{gap},\min}/E_{\mathrm{split},\max}$ over the twisted-boundary-condition torus. The minimum gap is maximized at $t_2=-0.35$ for all three interaction strengths. The ratio $E_{\mathrm{gap},\min}/E_{\mathrm{split},\max}$ can remain competitive, or become slightly larger, near $t_2=-0.37$, but the combined gap and momentum-space diagnostics select $t_2=-0.35$ as the representative point.

\clearpage

\section{Robustness against weak nearest-neighbor admixture}\label{sec:u1-robust}

\begin{figure}[htbp]
	\centering
	\includegraphics[width=0.82\textwidth]{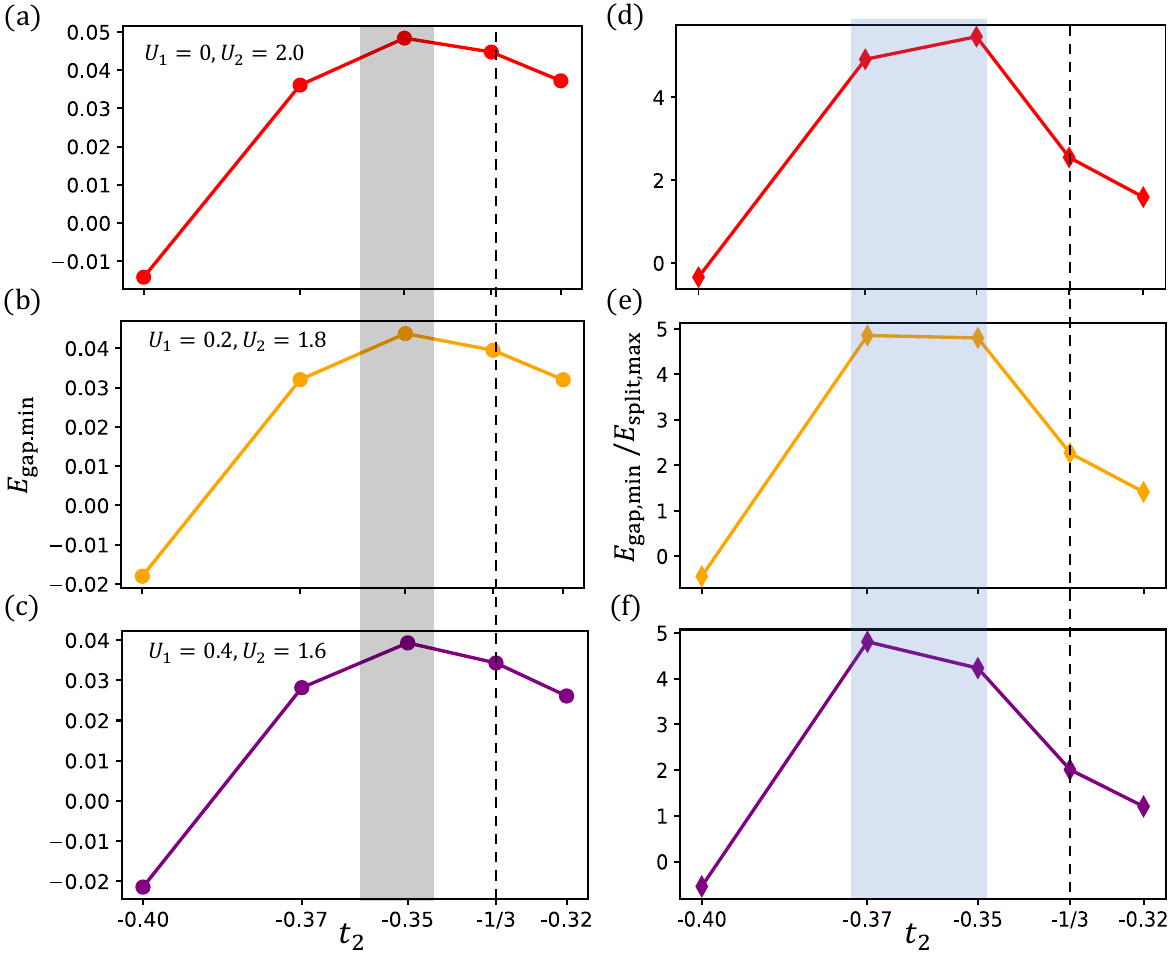}
	\caption{Robustness of the $\nu=1/5$ stability-enhanced window against weak nearest-neighbor interaction admixture. Panels (a)--(c) show the minimum many-body gap $E_{\mathrm{gap},\min}$ as a function of $t_2$ for $(U_1,U_2)=(0,2.0)$, $(0.2,1.8)$, and $(0.4,1.6)$, respectively. Panels (d)--(f) show the corresponding ratio $E_{\mathrm{gap},\min}/E_{\mathrm{split},\max}$ for the same interaction profiles. The gray marker in the gap panels highlights $t_2=-0.35$, where the minimum gap is maximal. In the ratio panels, the markers indicate the nearby candidate points $t_2=-0.37$ and $t_2=-0.35$.}
	\label{fig:u1-robust}
\end{figure}

We next test whether the enhancement is restricted to the pure third-neighbor interaction profile. To this end, we weakly admix nearest-neighbor repulsion while keeping the overall interaction scale approximately fixed. Figure~\ref{fig:u1-robust} compares $(U_1,U_2)=(0,2.0)$, $(0.2,1.8)$, and $(0.4,1.6)$, with $(0,2.0)$ serving as the common reference profile also shown in Fig.~\ref{fig:u2-robust}. The minimum gap is again maximized at $t_2=-0.35$ for all three profiles. As in the pure-$U_2$ comparison, the gap-to-splitting ratio may favor the neighboring point $t_2=-0.37$ for some profiles, but the gap maximum together with the momentum-space diagnostics in Fig.~\ref{fig:nk-sq-differ-fifth} selects $t_2=-0.35$ as the more uniform and robust representative point. This indicates that the $\nu=1/5$ enhancement is not tied to the fine-tuned choice $(U_1,U_2)=(0,2.0)$.

\clearpage

\end{document}